\documentclass[11pt,a4paper]{article}
\usepackage[utf8]{inputenc}
\usepackage{jheppub}
\usepackage{amsmath}
\usepackage{amsfonts}
\usepackage{amssymb}
\usepackage{graphicx} 
\usepackage{epsfig}
\usepackage{color}
\usepackage{amssymb}
\usepackage{bbold}
\usepackage{hhline}
\usepackage{tabu}
\usepackage{mathtools}

\newcommand{\ee}[1]{\begin{equation}#1\end{equation}}
\newcommand{\ea}[1]{\begin{align}#1\end{align}}

\providecommand{\f}[2]{\frac{{#1}}{{#2}}}

\title{Dark energy as a remnant of inflation and electroweak symmetry breaking}

\author[a]{Konstantinos Dimopoulos}
\author[b]{Tommi Markkanen}
\affiliation[a]{Consortium for Fundamental Physics, Physics Department,
	Lancaster University, Lancaster LA1 4YB, United Kingdom}
\affiliation[b]{Department of Physics, Imperial College London, SW7 2AZ, UK}

\abstract{It is shown that dark energy can be obtained from the interplay of the Higgs boson and the inflaton. A key element is the realization that electroweak symmetry breaking can trigger a {second phase of rolling} of the inflaton, which, when provided with the appropriate couplings between the fields, can be sufficiently slow to source accelerated expansion in the late time Universe. The observed dark energy density is obtained without fine-tuning of parameters or initial conditions due to an intricate conspiracy of numbers related to inflation, gravity and electroweak physics.}

\emailAdd{konst.dimopoulos@lancaster.ac.uk}
\emailAdd{t.markkanen@imperial.ac.uk}
\begin{document}
	\begin{flushleft}
		\hfill		  IMPERIAL/TP/2018/TM/03
	\end{flushleft}
	\maketitle	
\section{Introduction}

For explaining the observation of accelerated expansion of the Universe \cite{Riess:1998cb,Perlmutter:1998np}, 
dark energy is perhaps the most compelling suggestion. There is currently no experimental evidence that requires one to go beyond the simplest model, a non-zero cosmological constant $\Lambda$ in the gravitational Lagrangian \cite{Ade:2015rim}. From a theoretical point of view however a cosmological constant of the observed magnitude is highly problematic \cite{Martin:2012bt,Sola:2013gha}. Any quantum field gives rise to a vacuum energy contribution that is indistinguishable from $\Lambda$ but generically several orders of magnitude larger than what is required to match observations. This is the infamous "Cosmological Constant Problem", which in fact predates the discovery of dark energy. The standard "explaination" had been that some {\em unknown} mechanism sets the Cosmological Constant to zero. This is why, after the current acceleration was found, a wealth of explanations that go beyond strictly constant vacuum energy [6] have been put forward, all of which do not solve the Cosmological Constant problem but continue to rely to an unknown symmetry to explain $\Lambda$ away. Their aim is to account for the observed accelerated expansion, while assuming \mbox{$\Lambda=0$}. The majority of the proposed resolutions involve a slowly rolling scalar field often dubbed quintessence, which originated in \cite{Wetterich:1987fm,Peebles:1987ek,Ratra:1987rm,Caldwell:1997ii}. \footnote{Recently, a novel conjecture 	\cite{Obied:2018sgi} has suggested that quintessence is favoured over non-zero $\Lambda$, in order that the effective field theory is part of the string landscape and not the swampland \cite{Agrawal:2018own}.} Usually, the quintessence field is separate from the inflaton, which is believed to have sourced the spectrum of primordial perturbations. Models where dark energy and inflation are given by the one and the same scalar field do however exist \cite{Peebles:1998qn}, called quintessential inflation (see Ref.~\cite{Dimopoulos:2017zvq} for an updated list of references), and have some overlap with our proposal as does the symmetron mechanism \cite{Hinterbichler:2010es}. We also note a similarity to \cite{Burrage:2018dvt} analysing constraints on the Higgs sector from scalar-tensor 
theories.

\begin{figure}[ht]
	\begin{center}
		\includegraphics[width=0.6\textwidth,trim={0.3cm 0cm 0.1cm 0cm},clip]{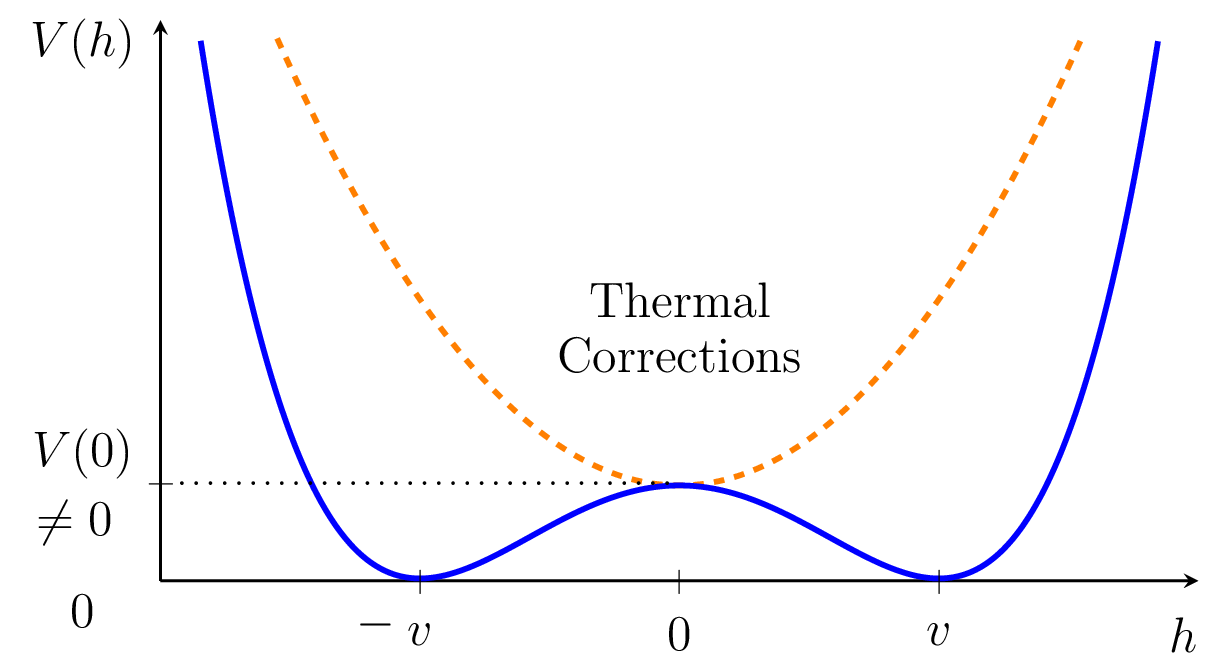}
	\end{center}
	\caption{The potential for the Higgs (\ref{eq:Hig}) with the choice $V(v)=0$. When $T>T_{\rm EW}$ thermal corrections will prevent spontaneous symmetry breaking and lead to a convex potential (the orange dashed curve). When $T<T_{\rm EW}$ the potential takes the mexican hat shape (the blue curve). 
		\label{fig:hig}}
\end{figure}

The Standard Model (SM) Higgs also introduces complications in regards to vacuum energy
 \cite{Martin:2012bt,Sola:2013gha}. Since a scalar degree of freedom of the SM Higgs doublet, 
$h$, gets a vacuum expectation value (VEV) at electroweak symmetry breaking the vacuum energy as given by the minimum of the Higgs potential $V(h_{\rm min})$ experiences a similar change: 
when the temperature of the Universe $T$ crosses the symmetry breaking threshold the thermally corrected potential (orange dashed curve in Fig.  \ref{fig:hig}\,) acquires a Mexican hat shape (blue curve in Fig.  \ref{fig:hig}\,) leading to a change in vacuum energy that is some $10^{55}$ times larger than the observed dark energy component. 
So even if in principle one could introduce an overall constant in the Lagrangian to explain the observed accelerated expansion this requires a tuning of 55 orders of magnitude.

As mentioned in the introduction, in dark energy model building it is generally assumed that for some unspecified reason, perhaps due to an underlying symmetry, the total vacuum energy vanishes in the broken phase. This gives the bare Higgs potential \ee{V(h)=\f{\lambda}{4}\left(h^2-v^2\right)^2\,,\label{eq:Hig}}
where $v$ is the VEV in the electroweak vacuum. With this assumption one is free to explain accelerated expansion effectively bypassing (some of) the issues that plague models with constant vacuum energy, but at the cost of introducing new degrees of freedom that often involve fine-tuned potentials and/or initial conditions. 

Here we will put forward a new approach for explaining dark energy. It requires no degrees of freedom beyond the SM Higgs $h$ and the inflaton $\phi$. 
The current phase of accelerated expansion is due to the {interplay} of the SM Higgs potential and the inflaton where the non-trivial dynamics at electroweak symmetry breaking via a direct coupling trigger a \textit{second phase of rolling} such that the inflaton may (again) source accelerated expansion.

In the following, we use natural units, where $c=\hbar=1$ and Newton's 
gravitational constant is \mbox{$8\pi G=M_{\rm P}^{-2}$}, with 
\mbox{$M_{\rm P}=2.43\times 10^{18}\,$GeV} being the reduced Planck mass.{ Throughout $h$ is the 1-point function or mean field.}

We parametrize the tree-level Higgs potential as in (\ref{eq:Hig}), and in terms of the temperature of the Universe  $T$ the value of the dressed potential at the minimum $V(h_{\rm min})$ i.e. vacuum energy has the following behaviour
\ee{V(h_{\rm min})=\begin{cases}V(0)=\f{\lambda}{4}v^4\,, &T>T_{\rm EW}\\V(v)=0\,, &T<T_{\rm EW}
\end{cases}\,.\label{eq:change}}
Strictly speaking, when $T\lesssim T_{\rm EW}$ the Higgs VEV is displaced from $h=v$ and hence does not result in $V(v)=0$ due to the non-zero thermal contribution. However, when the Universe cools below $T\sim T_{\rm EW}$ such corrections disappear quickly. 
For our purposes it will turn out to be sufficient to neglect all thermal corrections for $T<T_{\rm EW}$ and  describe the behaviour of $V(h)$ with (\ref{eq:change}). 

\section{The Conspiracy of Scales}
We can illustrate our mechanism by considering a model where the SM is modified to include the familiar Starobinsky model of inflation coming from a sizeable $R^2$ term \cite{Starobinsky:1979ty}. Writing explicitly only the gravitational sector and the Higgs potential the action then reads
\ee{S=\int \sqrt{-g}\bigg[\f{M_{\rm P}^2}{2}R+\f{1}{16\alpha^2}R^2-V(h)\bigg]\,,\label{eq:act}} where $\alpha$ is a dimensionless parameter fixed to give the observed spectrum of perturbations. 

After expressing the above in terms of a Lagrange multiplier field, a suitable Weyl scaling $g_{\mu\nu}\longrightarrow\Omega^{-2}g_{\mu\nu}$ 
and a field redefinition, we can extract the additional scalar degree of freedom, the inflaton $\phi$, introduced by $R^2$ leading to (see e.g. \cite{Kehagias:2013mya})
\ee{S=\int \sqrt{-g}\bigg[\f{M_{\rm P}^2}{2}R-\f{1}{2}(\partial_{\mu}\phi)^2-U(\phi,h)\bigg]\,,\label{eq:act0}}
where we have defined the combined potential
\ea{U(\phi,h)=\alpha^2M^4_{\rm P}\bigg(1-e^{-\sqrt{\f{2}{3}}\phi/M_{\rm P}}\bigg)^2 +e^{-\sqrt{\f{8}{3}}\phi/M_{\rm P}}V(h)\,.\label{eq:pot}}
Depending on the couplings of the Higgs, during inflation it behaves as a stochastic spectator with $V(h)\sim H^4\sim (\alpha M_{\rm P})^4$ \cite{Starobinsky:1994bd} or as a heavy field with no excitations $V(h)\sim 0$. Either way, since the primordial perturbations imply $\alpha\sim 10^{-5}$ from (\ref{eq:pot}) one may see that the presence of the Higgs has virtually no impact on the inflationary predictions. After inflation however, the coupling between $\phi$ and $h$ will turn out to be important.

\begin{figure}
	\begin{center}
		\includegraphics[width=0.6\textwidth,trim={0.3cm 0cm 0.1cm 0cm},clip]{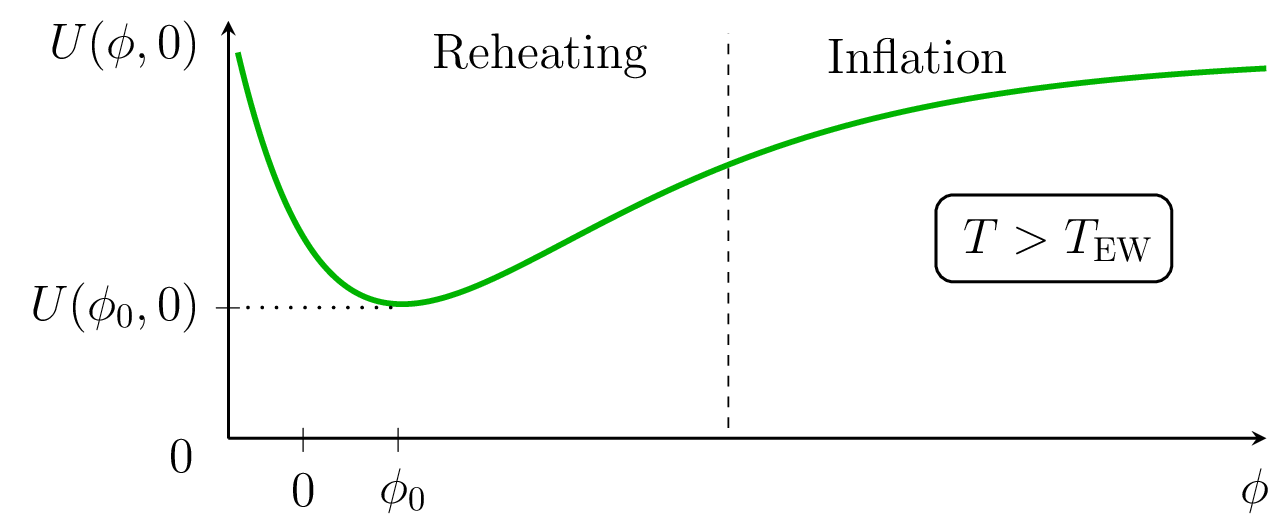}
				\includegraphics[width=0.6\textwidth,trim={0.3cm 0cm 0.1cm 0cm},clip]{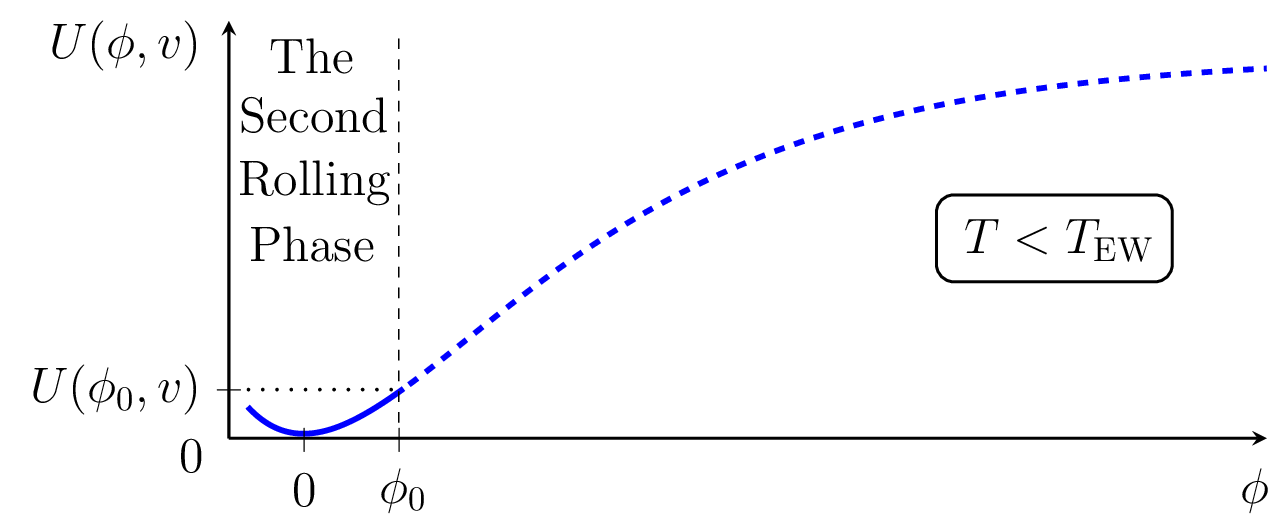}
	\end{center}
	\caption{The potential (\ref{eq:pot}) before electroweak symmetry breaking when $h=0$ (top) and after when $h=v$ (bottom). The solid blue curve is the region where the second rolling triggered by electroweak symmetry breaking takes place.\label{fig:sca}}
\end{figure}

When inflation has ended and reheating has taken place but $T\geq T_{\rm EW}$, the Higgs will reside in the origin $h=0$ and the inflaton in a vacuum at $\phi=\phi_0$ 
\ee{U'(\phi_{0},0)=0\quad \Rightarrow\quad e^{\sqrt{\f{2}{3}}\phi_{0}/M_{\rm P}}={1+\f{V(0)}{\alpha^2M^4_{\rm P}}}\label{eq:min}\,.}
Since $V(0)/(\alpha^2M_{\rm P}^4)$ is very small but non-zero, the vacuum for the inflaton will be ever so slightly displaced from $\phi=0$ due to the non-vanishing $V(0)$ as shown by the green curve in Fig. \ref{fig:sca}\,. This results in the usual non-zero electroweak vacuum energy contribution $U(\phi_{0},0)\approx V(0)=({\lambda}/{4})v^4$, as can easily be calculated from (\ref{eq:pot}) and (\ref{eq:min}). Although this is a large value compared to the energy density today it makes little difference when the temperature of the Universe is $T_{\rm EW}$ or higher\footnote{$V(0)/(\f{\pi^2}{30}g^* T_{\rm EW}^4)\sim 10^{-3}$ for $g^*\sim 100$.}.

However, when $T\leq T_{\rm EW}$ the Higgs field acquires a VEV, $h=v$, and its vacuum energy vanishes, $V(h_{\rm min})=0$, triggering the rolling of the inflaton towards the new minimum at $\phi=0$ shown by the solid blue curve in Fig. \ref{fig:sca}\,. The initial value for the rolling is $\phi_{0}$, which leads to
\ee{U(\phi_{0},v)\approx \f{V^2(0)}{\alpha^2 M_{\rm P}^4}=\f{\lambda^2 v^8}{16\alpha^2 M_{\rm P}^4}\,,\label{eq:CC}}
including only the leading term in $V(0)/(\alpha^2 M_{\rm P}^4)$.
Using the best-fit values for {$v=246\text{GeV}$, $\lambda=0.129$ \cite{deFlorian:2016spz} 
and $\alpha = 9.97\cdot10^{-6}$} corresponding to the observed spectrum of perturbations for $R^2$-inflation \cite{Ade:2015lrj}, one sees a surprising and a remarkable numerical conspiracy: 
\ee{\f{\lambda^2 v^8}{16\alpha^2 M_{\rm P}^4}\approx 4.0\times 10^{-48}\text{GeV}^4\,,\label{eq:cc2}}
which is very close to the energy density of dark energy as observed today, $\rho_\Lambda\sim2.58\times 10^{-47}\text{GeV}^4$ \cite{Ade:2015xua}. This is the first of the two novel findings of this article. We stress that no parameter was tuned in order to match the correct magnitude. All inputs, including the initial condition $\phi_0$, are fixed by observations related to inflation, gravity and electroweak physics. {It is well-known that the electroweak scale $\sim v$ is roughly the geometric mean of $M_{\rm P}$ and $\rho_\Lambda\!\!\,^{1/4}$, however in terms of the energy densities this is still off by a factor of $\rho_\Lambda(v^8/M_{\rm P}^4)^{-1}\sim10^8$, as oppose to (\ref{eq:cc2}) which gives $\rho_\Lambda$ up to a factor of $\sim 6$.}  {If $\phi$ rolls slowly enough}, the second rolling phase can result in an effective vacuum energy contribution, precisely as in quintessence models.

Unfortunately, the potential (\ref{eq:pot}) with $h=v$ does not lead to a slowly rolling $\phi$ after electroweak symmetry breaking. This is evident from the effective mass of $\phi$
\ee{U''(\phi_{0},v)\equiv m_{\rm eff}^2\sim(\alpha M_{\rm P})^2\sim (10^{13}\text{GeV})^2\,,\label{eq:meff}}
which clearly is far too large a value: the field will roll sufficiently slowly only when 
the Hubble friction dominates over the effective mass, $H\gg m_{\rm eff}$. 
Therefore, almost instantly after the electroweak phase transition the inflaton will roll away from $\phi=\phi_0$ (or decay completely). 
\section{The Bait-and-switch Mechanism}
The second important discovery of this work is that with a simple modification $\phi$ can be made light after electroweak symmetry breaking such that it remains frozen until, $H\sim H_{0}\sim m_{\rm eff}\sim10^{-42}$GeV, where $H_0$ is the Hubble constant today. Furthermore, this can be achieved without affecting the crucial relation leading to the correct magnitude of dark energy (\ref{eq:CC}).

We now make use of a non-canonical kinetic term 
\ee{(\partial_\mu \phi)^2\quad\longrightarrow\quad\bigg(b\f{M_{\rm P}}{\phi}\bigg)^2(\partial_\mu \phi)^2\equiv (\partial_\mu \chi)^2\,,\label{eq:alpha}}
where $b$ is a dimensionless parameter of $\mathcal{O}(1)$. Due to the pole at $\phi=0$ the dynamics become modified such that $\phi$ will never be able to roll to the origin. In the canonical variable $\chi$ this appears as a very flat potential at large negative values of $\chi$. 

{This feature that increasing the prefactor in front of a canonical kinetic term generically leads to slower rolling down the potential was first exploited in the context of quintessence in Refs. \cite{Hebecker:2000au,Hebecker:2000zb}. Such non-canonical kinetic terms have a theoretical motivation in models with  radiatively induced breaking of conformal symmetry as discussed in Ref. \cite{Wetterich:2002wm}. We also point out that similar non-canonical kinetic terms appear in the supergravity framework and have recently attracted significant interest in the context of $\alpha$-attractors \cite{Kallosh:2013tua,Kallosh:2013yoa}. The goal of this work is however not finding a working top-down theory. Rather, we provide a demonstration that the construction of phenomenological models with the desired behaviour at early and late times is possible, in various frameworks and without fine-tuning.}

\begin{figure}
	\begin{center}
		\includegraphics[width=0.6\textwidth,trim={0.3cm 0cm 0.1cm 0cm},clip]{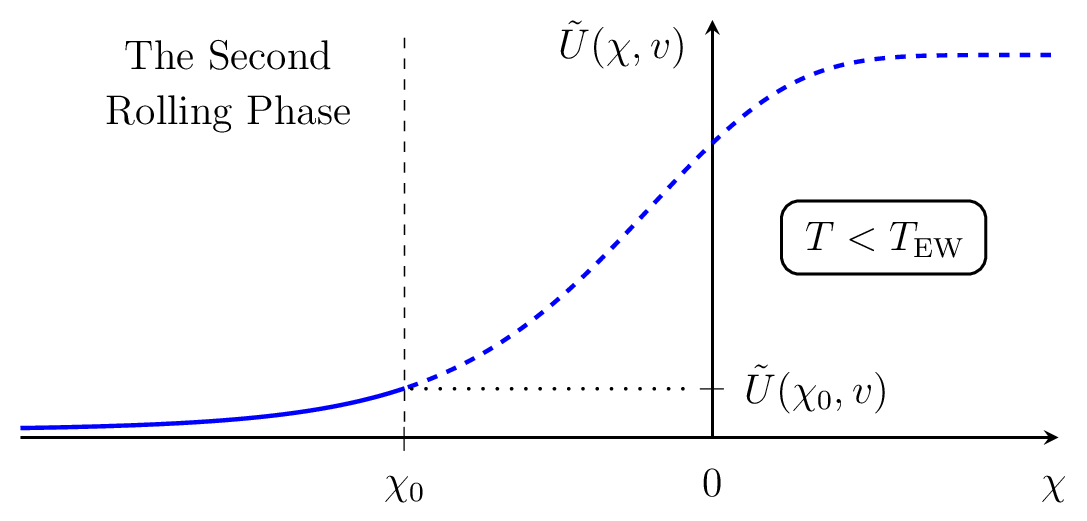}
	\end{center}
	\caption{The potential (\ref{eq:pot}) with $h=v$ with the kinetic term (\ref{eq:alpha}) in the canonical variable $\chi$.\label{fig:sca2}}
\end{figure}

The action (\ref{eq:act0}) with $h=v$, but with the kinetic term (\ref{eq:alpha}), in terms of $\chi$ has the potential 
\ee{U\left(\phi(\chi),v\right)\equiv\tilde{U}(\chi,v)\,,} which is plotted in Fig. \ref{fig:sca2}\,. 
In a way one may think of the solid blue curve  in Fig. \ref{fig:sca2}\, as being a stretched version of the solid blue in curve in Fig. \ref{fig:sca}\,, $[0,\phi_{0}]\rightarrow [-\infty,\chi_{0}]$. Importantly, the slope of the potential in terms of $\chi$ is significantly different than in terms of $\phi(\chi)$ but its value at some $\phi(\chi)$ is not. Solving (\ref{eq:alpha}) we can express the minimum $\phi_{0}$ (\ref{eq:min}) with $\chi_0$ to leading order in $V(0)/(\alpha^2 M_{\rm P}^4)$
\ee{{\phi(\chi)}=M_{\rm P}e^{\f{\chi}{b M_{\rm P}}},\quad\Rightarrow\quad {\chi_{0}}={b}M_{\rm P}\log\bigg(\f{\sqrt{{3}}V(0)}{\sqrt{2}\alpha^2 M_{\rm P}^4}\bigg)\,.\label{eq:min2}}
Close to $\phi=0$ the potential (\ref{eq:pot}) can be approximated with $U(\phi,v)\approx \frac{2}{3}\alpha^2 M_{\rm P}^2\phi^2$, which as a function of $\chi$ is
\ee{\tilde{U}(\chi,v)=\f{2}{3}\alpha^2M_{\rm P}^4e^{\f{2\chi}{b M_{\rm P}}}\,.\label{eq:chot}}
The potential (\ref{eq:chot}) straightforwardly gives the two important relations describing the dynamics of $\chi$
\ee{\tilde{U}(\chi_{0},v)=\f{V^2(0)}{\alpha^2M_{\rm P}^4}\,;\qquad \tilde{U}''(\chi_{0},v)=\f{4}{(b M_{\rm P})^2} \f{V^2(0)}{\alpha^2M_{\rm P}^4}\,.}
As one may see from (\ref{eq:CC}) and (\ref{eq:cc2}) the above gives $\tilde{U}(\chi_{0},v)\sim\rho_\Lambda\sim M_{\rm P}^2H_{0}^2$  and $\tilde{U}''(\chi_{0},v)\sim {M_{\rm P}^{-2}}\rho_\Lambda\sim H_{0}^2$ where $\rho_\Lambda$ is again the observed dark energy density. This means that if $\phi=\phi_0$ at electroweak symmetry breaking the inflaton will stay frozen until $H\sim H_{0}$ and hence with the form of the kinetic term in (\ref{eq:alpha}), we have the behaviour of thawing quintessence \cite{Caldwell:2005tm}.

As shown, a model with a non-canonical kinetic term (\ref{eq:alpha}) and the potential (\ref{eq:pot}) has flat plateaus that support inflation for $\chi\rightarrow\infty$ and dark energy for $\chi\lesssim\chi_0$, which is illustrated in Fig. \ref{fig:sca2}\,. It is unfortunately not apparent that starting from the inflationary plateau the field has the appropriate dynamics that lead to the correct behaviour at late times: the minimum (\ref{eq:min2}), which the field must reach in order to provide the correct value for dark energy, is very far away from the origin, $\chi_0\sim -b\cdot130 M_{\rm P}$. Furthermore, the end of inflation for the potential (\ref{eq:pot}) occurs at $\phi\sim M_{\rm P}$ $\Rightarrow \chi\sim 0$. Although the precise way in which the field behaves depends on the cosmological equation of state after inflation, since the effect of the non-canonical kinetic term is to make $\chi$ light it generally will not be possible for it to get to $\chi_0$ from the origin before being stopped by Hubble friction. We can conclude that despite providing the correct dynamics from $\chi_0$ onwards, having a non-canonical kinetic term present throughout the evolution is problematic. 


One way to get around the above issue is switching on the non-canonical kinetic term (\ref{eq:alpha}) after the field has reached the minimum (\ref{eq:min}) but before electroweak symmetry breaking. This we name the \textit{bait-and-switch} mechanism: first the field rolls unhindered to the temporary minimum at $\phi=\phi_0$, after which the potential is stretched such that once the minimum disappears the field remains frozen. There are likely many scenarios giving rise to such behaviour, but here we opt for the natural choice where the appearance of the non-canonical kinetic term is triggered by the electroweak phase transition.

{We may now proceed to write an example of an action in the mean field approximation with the correct prediction for dark energy as well as inflation, containing only the inflaton and the Higgs and without introducing fine-tuned parameters 
\ea{S&=\int \sqrt{-g}\bigg\{\f{M_{\rm P}^2}{2}R-\f{1}{2}\bigg[1+\bigg(\f{h}{v}\bigg)^2\bigg(b\f{M_{\rm P}}{\phi}\bigg)^2\bigg](\partial_\mu \phi)^2\nonumber \\&-\alpha^2M^4_{\rm P}\bigg(1-e^{-\sqrt{\f{2}{3}}\phi/M_{\rm P}}\bigg)^2 -c\bigg(1-e^{-\sqrt{\f{8}{3}}\phi/M_{\rm P}}\bigg)V(h)-V(h)\bigg\}\,.\label{eq:act2}}
}In the above $(h/v)^2$ is the important factor triggering the flattening of the potential after electroweak symmetry breaking. The $b$ and $c$ parameters are introduced to match the theory to current observations of dark energy. Importantly, they are $\mathcal{O}(1)$. According to the analysis of \cite{Dimopoulos:2017tud} $b\gtrsim3$ evades the bounds from the Planck satellite observations \cite{Ade:2015xua}, while setting $c\approx2.5$ leads to agreement with the observed $\rho_\Lambda$. Hence the introduced terms are not fine-tuned, as advertised. All other parameters come from observations. 

During inflation the non-canonical kinetic term makes $h$ heavy, 
as can be shown with the slow-roll expansion
\ee{\bigg(\f{h}{v}\bigg)^2\bigg(b\f{M_{\rm P}}{\phi}\bigg)^2(\partial_\mu \phi)^2\sim \f{\epsilon H^2 }{v^2}h^2 M_{\rm P}^2\gg h^2 M_{\rm P}^2\,,}
where $\epsilon =-\dot{H}/H^{2}$ is the slow-roll parameter. This suppresses the fluctuations in the Higgs such that to a very good approximation $h=0$ and the inflationary predictions are identical to those suggested by (\ref{eq:act0}). Until successful reheating this contribution continues to make the Higgs effectively heavy. 

{Going beyond the mean field approximation and taking the action (\ref{eq:act2}) at face value we see that the SM phenomenology of today will be affected by the mass dimension 6 operator $\sim(\f{h}{v})^2(\partial \chi)^2$, which is suppressed only by the electroweak scale. It is however possible to write a different theory which is identical to (\ref{eq:act2}) at the mean field level but where such an interaction is not generated. 
Instead of the kinetic interaction terms in (\ref{eq:act2}) we may use ${M_{\rm P}^2b^2}\{\partial_\mu[{h}/{v}\log({\phi}/{M_{\rm P}})]\}^2$, 
which leads to a suppression of interactions between the inflaton and the Higgs  $\sim\exp\left\{{\chi_0}/({b M_{\rm P})}\right\}\sim10^{-56}$, as one can show with
the change of variables $\tilde{\chi}=(h/v){\chi}$ for which 
\ee{(\partial_\mu \phi)^2+{M_{\rm P}^2b^2}\bigg\{\partial_\mu\bigg[\f{h}{v}\log\bigg(\f{\phi}{M_{\rm P}}\bigg)\bigg]\bigg\}^2=M_{\rm P}^2\bigg[\partial_\mu e^{\f{\tilde{\chi}}{b(h/v)M_{\rm P}}}\bigg]^2+(\partial_{\mu}\tilde{\chi})^2\label{eq:kin3}}
and expanding around the VEVs $\tilde{\chi}=\chi_0$ and $h=v$.
	
Couplings to other SM fields (and beyond) not visible in (\ref{eq:act2}) may be required by a particular reheating setup or possibly by radiative corrections, however they are not likely to spoil  our mechanism. For example, before electroweak symmetry breaking a term $\sim\phi\bar\psi\psi$ would simply give rise to a faster decay into the vacuum at $\phi_0$ (green curve in Fig. \ref{fig:sca}), but after the non-canonical kinetic term is triggered it becomes exponentially suppressed, much like the interactions with the Higgs.
The same applies for other interactions and also allows the evasion of the 5th force problem, which typically plagues quintessence model building \cite{Carroll:1998zi}. This arises generically from flattening due to a non-canonical kinetic term \cite{Kallosh:2016gqp}.}

The behaviour in the late time Universe is quite insensitive to the specifics of reheating as well as the precise dynamics at the triggering of the second phase of rolling: perturbations $\Delta \chi \lesssim M_{\rm P}$ around the minimum (\ref{eq:min2}) have only a small effect on the predictions. By solving the equation of motion for $\chi$ it can further be shown that at $T=T_{\rm EW}$ initial kinetic energies $\dot{\chi}\sim T_{\rm EW}^2$ dissipate before matter domination giving a displacement of at most $\Delta\chi\lesssim (T_{\rm EW}/H_{\rm EW})T_{\rm EW}\sim M_{\rm P}$ at current scales\footnote{For $a\propto\sqrt{t}$, $\dot{\chi}(t_0)=T_0^2$ and ${\chi}(t_0)=0$, $\Box \chi=0$ gives $\chi \sim \big(1-(t_0/t)^{1/2}\big)t_0 T_0^2$ and $\dot{\chi}^2\sim (t_0/t)^3T_0^4\sim T_0^{-2}T^6$.}.

{The model in (\ref{eq:act2}) can, of course, no longer considered to be the pure Starobinsky model of inflation (\ref{eq:act}). Nonetheless it is an example of a phenomenological model that correctly reproduces the inflationary epoch as well as the dark energy dominated phase with no fine-tuned parameters. The way we arrived at it suggests a more general strategy with which one may find other working examples: starting from a model of inflation, by introducing an $\mathcal{O}(1)$ coupling between the inflaton and the electroweak Higgs potential and introducing a term that is negligible during inflation but halts the rolling of the field after electroweak symmetry breaking all the required qualitative features, namely the conspiracy of scales (\ref{eq:cc2}) and the bait-and-switch mechanism, can be realized. 

As an example}, suppose that the potential for the inflaton for small $\phi/M_{\rm P}$ can be approximated with only a quadratic term and that the leading interaction between $\phi$ and $h$ is of the form $\sim\phi V(h)$
\ea{U(\phi,h)=\f{1}{2}m^2\phi^2+V(h)+\tilde{c}\f{\phi}{M_{\rm P}}V(h)\,,\label{eq:pot3}}
where $\tilde{c}$ is again a dimensionless coupling. Precisely as (\ref{eq:pot}) the above will after inflation trap $\phi$ at $\phi_0\neq0$ until electroweak symmetry breaking which triggers a second rolling towards the origin. More or less the identical steps that lead to Eq. (\ref{eq:CC}) give for the model in Eq. (\ref{eq:pot3})
\ee{U(\phi_{0},v)= \f{\tilde{c}^2V^2(0)}{2m^2 M_{\rm P}^2}\,.\label{eq:CC2}}
The value for $m$ that leads to the correct size of primordial perturbations for quadratic inflation $m\sim 6\cdot10^{-6}M_{\rm P}$, and $|\tilde{c}|\sim\mathcal{O}(1)$ result in a $U(\phi_{0},v)$ that coincides with $\rho_\Lambda$ demonstrating {that generically the inflaton and Higgs conspire to result in potential energy $\sim\rho_\Lambda$ at electroweak symmetry breaking. {Although the quadratic model of inflation is currenty under tension we emphasize that eq. (\ref{eq:pot3}) does not need to be true during inflation for the model to work, which implies that there are likely other inflationary potentials that lead to successful models.}
	
Finally, we present a gravitational realization of the bait-and-switch mechanism. It comes with a kinetic term of the form
\ee{\bigg[1+\tilde{b}^2\bigg(\f{R^2}{v^4}+\f{\phi}{M_{\rm P}}\bigg)^{-2}\bigg](\partial_\mu\phi)^2\,,\label{eq:bs2}}
where again $\tilde{b}\gtrsim3$. Roughly, one can use any scale above $v$ but below the reheating scale for the desired result\footnote{Note that $R$ is never 0 due to the conformal anomaly.} and furthermore one could just as well instead of $R^2$ make use of $R_{\mu\nu}R^{\mu\nu}$ or $R_{\mu\nu\alpha\beta}R^{\mu\nu\alpha\beta}$.}
\section{Conclusions}
{From a purely phenomenological point of view we have demonstrated how couplings of $\mathcal{O}(1)$ between the inflaton $\phi$ and the Higgs $h$ can give rise to dynamics where electroweak symmetry breaking triggers a second rolling phase of the inflaton, which is slow enough to manifest as dark energy, in agreement with observations. Although such couplings can partly have a motivation in the context of $f(R)$ theories as implied by the action (\ref{eq:act0}) the models we have presented should not be viewed as fundamental. Rather, our focus was to highlight that the class of models possessing the desired features is broad. This evident from the different forms of potentials in (\ref{eq:act2}) and (\ref{eq:pot3}) and furthermore in the three distinct non-canonical kinetic terms in (\ref{eq:act2}), (\ref{eq:kin3}) and (\ref{eq:bs2}). The predictions are robust in the sense that they are not sensitive to the specifics of reheating or additional couplings to the SM. The usual issues regarding 5th forces are not present. Since we have not attempted building working top down constructions, finding a successful first principle theory is required in order for our mechanism to be a viable explanation for the observed accelerated acceleration.} 

Our approach avoids fine-tuning because of relation (\ref{eq:CC}) (or (\ref{eq:CC2})), which is perhaps the most surprising discovery of this work. This realization that up to a factor of $\sim 6$ the observed dark energy density is expressible with parameters describing inflation, gravity and electroweak physics may be more than just a coincidence, potentially providing an important clue of the nature of dark energy.
\acknowledgments{We thank Malcolm Fairbairn and Tomohiro Fujita for probing questions and Claudia de Rham for valuable comments on the draft. KD is supported (in part) by the Lancaster-Manchester-Sheffield Consortium for Fundamental Physics under STFC grant: ST/L000520/1 and TM is supported by the STFC grant ST/P000762/1.}
\bibliography{B_S.bib}
\end{document}